\documentclass[preprint,12pt, aps,amsmath,amssymb,nofootinbib,superscriptaddress,hyperref,compact]{revtex4} 
\usepackage[utf8]{inputenc}
\usepackage{epsfig} 
\usepackage{wasysym} 
\usepackage{mathrsfs} 
\usepackage{graphicx} 
\usepackage{amsfonts} 
\usepackage{amsbsy} 
\usepackage{amscd} 
\usepackage{pstricks} 
\usepackage{multirow}
\usepackage{slashed} 
\usepackage{tikz}
\usetikzlibrary{decorations.pathmorphing,decorations.markings}
\newcommand{\beq}{\begin{equation}}
\newcommand{\eeq}{\end{equation}}
\newcommand{\bea}{\begin{eqnarray}}
\newcommand{\eea}{\end{eqnarray}}
\newcommand{\beas}{\begin{eqnarray*}}
\newcommand{\eeas}{\end{eqnarray*}}
\newcommand{\bi}{\begin{itemize}}
\newcommand{\ei}{\end{itemize}}

\def\tev{\,{\ifmmode\mathrm {TeV}\else TeV\fi}}
\def\gev{\,{\ifmmode\mathrm {GeV}\else GeV\fi}}
\def\to{\rightarrow}

\allowdisplaybreaks

\begin{document}

\title{
Associated Higgs boson production with $Z$ boson 
in the minimal $U(1)_X$ extended Standard Model
}

\author{Arindam Das\footnote{arindamdas@oia.hokudai.ac.jp}}
\affiliation{Department of Physics, Osaka University, Toyonaka, Osaka 560-0043, Japan}
\affiliation{Institute for the Advancement of Higher Education, Hokkaido University, Sapporo 060-0817, Japan}
\affiliation{Department of Physics, Hokkaido University, Sapporo 060-0810, Japan}
\author{Nobuchika Okada\footnote{okadan@ua.edu}}
\affiliation{Department of Physics and Astronomy, 
University of Alabama, \\ 
Tuscaloosa,  Alabama 35487, USA 
}
\preprint{\textbf{OU-HET-1068}}
\begin{abstract}
The minimal $U(1)_X$ extension of the Standard Model (SM) is a simple and well-motivated extension of the SM,  
  which supplements the SM with the seesaw mechanism for naturally generating the light neutrino masses 
  and offers various interesting phenomenologies. 
In the model, the $U(1)_X$ charge of each SM field is characterized 
  by the $U(1)_X$ charge of the SM Higgs doublet with a free parameter $x_H$.  
Due to the $U(1)_X$ charge of the Higgs doublet, the Higgs boson has a trilinear coupling 
  with the $Z$ boson and the $U(1)_X$ gauge boson ($Z^\prime$).   
With this coupling, a new process for the associated Higgs boson production 
  with a $Z$ boson arises through a $Z^\prime$ boson in the $s$-channel at high energy colliders. 
In this paper, we calculate the associated Higgs boson production at high energy colliders
  and show interesting effects of the new $Z^\prime$ boson mediated process, 
  which can be tested in the future.  

\end{abstract}

\maketitle

\section{Introduction} 

The discovery of the Higgs boson \cite{Chatrchyan:2012xdj,Aad:2012tfa} of the Standard Model (SM) 
  is definitely a very important milestone in the history of particle physics. 
It ends the quest of searching for the last SM particle after a long time and then further solidifies the plinth of the SM. 
Although the observation of the Higgs boson ensures its existence and its role for the electroweak symmetry breaking, 
  more precise measurements of the Higgs boson properties are necessary for verifying the SM Higgs sector 
  and/or for revealing new physics beyond the SM (BSM).

A simple extension of the SM can be possible by an additional $U(1)$ gauge symmetry \cite{Appelquist:2002mw} 
  which predicts a neutral BSM gauge boson ($Z^\prime$). 
This extra gauge boson has been searched for at the Tevatron \cite{Aaltonen:2008vx, Abazov:2010ti}, LEP \cite{Chiappetta:1996km,Barger:1996kr,Lynch:2000md} and LHC \cite{Aad:2019fac,CMS:2019tbu} with final state dileptons/dijes. 
With the precision measurements by the LEP experiment, the strong constraints on the $Z-Z^\prime$ mixing 
  has been studied in \cite{Carena:2004xs,Petriello:2008pu}. 
The collider phenomenology of the $Z^\prime$ boson has been intensively studied (see, for example,  
\cite{Rizzo:1985kn,Nandi:1986rg,Baer:1987eb,Barger:1987xw,Gunion:1987jd,Hewett:1988xc,Barger:1988sq,Feldman:2006wb,Langacker:2008yv,Lee:2008cn,Salvioni:2009mt,Barger:2009xg,Li:2013ava,Banerjee:2015hoa,Deppisch:2019ldi}) 
  with the final state leptons and quarks. 
Apart from the pair fermion productions, another important process to study the $Z^\prime$ boson 
  is an associated production of the SM Higgs boson ($h$) with the SM $Z$ boson, 
  such as $pp\to Z^\prime \to Z h$ at the LHC,  
  which can arise from a trilinear coupling $Z^\prime-Z-h$ if the SM Higgs doublet has an extra $U(1)$ gauge charge. 

To study the associated Higgs boson production, we consider in this paper 
  the minimal $U(1)_X$ extension of the SM \cite{Oda:2015gna,Das:2016zue},  
  where the particle content is the same as the minimal $B-L$ (baryon number minus lepton number) model \cite{Davidson:1987mh,Davidson:1980br, Davidson:1983fe, Mohapatra:1980qe,Marshak:1979fm,Wetterich:1981bx,Masiero:1982fi,Buchmuller:1991ce} 
  while the $U(1)_X$ charge of each SM particle is defined as a linear combination of its $U(1)_Y$ hyper-charge and $U(1)_{B-L}$ charge. 
In the presence of three right-handed neutrinos (RHSs), this model is free from all the gauge and the mixed gauge-gravitational anomalies. 
After the breaking of $U(1)_X$ and electroweak symmetries, the seesaw mechanism works to generate light neutrino masses naturally. 
The minimal $U(1)_X$ model provides interesting high energy collider phenomenologies. 
For the current LHC bound from the search for the $Z^\prime$ boson resonance with dilepton final states, 
  see, for example, \cite{Okada:2016tci, Okada:2017dqs, Okada:2018tgy}. 
A suitable choice of the $U(1)_X$ charges leads to a significant enhancement of the RHN pair production cross section 
   from the $Z^\prime$ boson which can increase the potential of the RHN discovery at the future high energy colliders 
   through the prompt RHN decay \cite{Das:2017flq,Das:2017deo,Das:2019pua,Deppisch:2019kvs,Choudhury:2020cpm} 
   and the displaced vertex search for a long-lived  RHN \cite{Deppisch:2018eth,Das:2018tbd,Das:2019fee,Chiang:2019ajm}. 
The main point of the present paper is to show interesting effects of the $Z^\prime$ mediated process 
   on the associated Higgs boson production at high energy colliders. 
   
We arrange this paper as follows: In Sec.~\ref{Sec2} we describe the minimal $U(1)_X$ model and the interactions relevant to our analysis. 
In Sec.~\ref{Sec3} we calculate the $Z h$ production cross sections at the proton-proton, electron-positron and muon colliders where we 
show the effects of $Z^\prime$ boson mediated process. 
We conclude this paper in Sec.~\ref{Sec4}.
   
\section{Minimal gauged $U(1)_X$ extension of the Standard Model}
\label{Sec2}
\begin{table}[t]
\begin{center}
\begin{tabular}{|c|ccc|c|}
\hline
      &  SU(3)$_c$  & SU(2)$_L$ & U(1)$_Y$ & U(1)$_X$  \\ 
\hline
$q^{\alpha}_{L}$ & {\bf 3 }    &  {\bf 2}         & $ 1/6$       & $(1/6) x_{H} + (1/3) x_{\Phi}$    \\
$u^{\alpha}_{R}$ & {\bf 3 }    &  {\bf 1}         & $ 2/3$       & $(2/3) x_{H} + (1/3) x_{\Phi}$  \\
$d^{\alpha}_{R}$ & {\bf 3 }    &  {\bf 1}         & $-1/3$       & $-(1/3) x_{H} + (1/3) x_{\Phi}$  \\
\hline
$\ell^{\alpha}_{L}$ & {\bf 1 }    &  {\bf 2}         & $-1/2$       & $(-1/2) x_{H} - x_{\Phi}$  \\
$e^{\alpha}_{R}$    & {\bf 1 }    &  {\bf 1}         & $-1$                   & $-x_{H} - x_{\Phi}$  \\
\hline
$H$            & {\bf 1 }    &  {\bf 2}         & $- 1/2$       & $(-1/2) x_{H}$  \\  
\hline
$N^{\alpha}_{R}$    & {\bf 1 }    &  {\bf 1}         &$0$                    & $- x_{\Phi}$   \\

$\Phi$            & {\bf 1 }       &  {\bf 1}       &$ 0$                  & $ + 2x_{\Phi}$  \\ 
\hline
\end{tabular}
\end{center}
\caption{
Particle content of  the minimal $U(1)_X$ model, 
  where $x_H$ and $x_\Phi$ are real parameters, and $\alpha=1, 2, 3$ is the three generation flavor index. 
Since the $U(1)_X$ gauge coupling is a free parameter, we fix $x_\Phi=1$ without loss of generality. 
}
\label{tab1}
\end{table}  

The simple gauged $U(1)$ extension of the SM is based on the gauge group $SU(3)_c\otimes SU(2)_L\otimes U(1)_Y\otimes U(1)_X$. 
The particle content of the model is listed in Tab.~\ref{tab1}. 
We consider a conventional case that the $U(1)_X$ charge assignment is flavor-universal \cite{Appelquist:2002mw}. 
In addition to the $U(1)_X$ gauge boson $Z^\prime$, three RHNs and one $U(1)_X$ Higgs field ($\Phi$) are introduced.  
The particle content of the model is basically the same as the one of the minimal $B-L$ model 
  while the $U(1)_X$ charge of each SM field is generalized to a linear combination of 
  its $U(1)_Y$ and $B-L$ charges with real parameters $x_H$ and $x_\Phi$. 
In the presence of three  RHNs, this model is anomaly-free.   
Since the $U(1)_X$ gauge coupling is a free parameter, we can set $x_\Phi=1$ without loss of generality. 
The minimal $B-L$ model is realized in the minimal $U(1)_X$ model as the limit of $x_H \to 0$.

With the RHNs, the extended part of the SM Yukawa sector can be written as 
\bea
\mathcal{L} _{Y}\supset -\sum_{\alpha, \beta=1}^{3} Y_{D}^{\alpha \beta} \overline{\ell_{L}^{\alpha}} H N_{R}^{\beta} 
  -\frac{1}{2} \sum_{\alpha=1}^{3} Y_{N}^{\alpha} \Phi \overline{N_{R}^{\alpha \, C}} N_{R}^{\alpha}+ \rm{H. c.}, 
\label{U1XYukawa}
\eea 
where $C$ stands for the charge-conjugation. 
The first and second terms in the right-hand side are the Dirac-type and Majorana-type Yukawa interactions, respectively. 
Here, we work in the bases where $Y_N$ is flavor-diagonal. 
After the $U(1)_X$ symmetry breaking by a vacuum expectation value (VEV) of the Higgs field $\Phi$, 
   the Majorana masses for the RHNs are generated, 
   while the neutrino Dirac mass terms are generated from the SM Higgs doublet VEV. 
With these mass terms, the seesaw mechanism works to naturally generate the light SM neutrino masses.

The (renormalizable) Higgs potential of the model is given by 
\begin{align}
  V \ = m_\Phi^2 (\Phi^\dag \Phi)+\lambda_\Phi (\Phi^\dag \Phi)^2+ \ m_H^2(H^\dag H)+\lambda_H (H^\dag H)^2+ \lambda^\prime (H^\dag H)(\Phi^\dag \Phi) \, .
  \label{pot1}
  \end{align}
For simplicity, we assume $\lambda^\prime$ is negligibly small and hence 
   the scalar potential can be analyzed separately for $\Phi$ and $H$ as a good approximation.  
By choosing the parameters suitably, 
  we can realize the $U(1)_X$ and electroweak symmetry breaking at the potential minimum 
  and parametrize the Higgs fields in the unitary gauge as follows: 
 \begin{align}
  \quad 
 \Phi \ = \  \frac{1}{\sqrt{2}}(v_\Phi + \phi)  \,   \quad {\rm and} \, \quad
 H  \ = \ \frac{1}{\sqrt{2}}\begin{pmatrix} v+h\\0 
  \end{pmatrix} \, ,
  \end{align}
where $v_\Phi$ and $v \simeq 246$ GeV are the Higgs VEVs, 
  and $\phi$ and $h$ are physical Higgs bosons. 
After the symmetry breaking, the mass of $Z^\prime$ boson, 
  the Majorana masses for the RHNs and the neutrino Dirac masses are generated as
$m_{Z^\prime}=g_X \sqrt{4 v_\Phi^2+  \frac{1}{4}x_H^2 v^2} \simeq \ 2 g_X v_\Phi \quad (v_\Phi \gg v)$
$m_{N_i}= \frac{Y^i_{N}}{\sqrt{2}} v_\Phi$ and $m_{D}^{ij}= \frac{Y_{D}^{ij}}{\sqrt{2}} v$, respectively, 
where $g_X$ is the $U(1)_X$ gauge coupling. 
Here, we have used the LEP~\cite{LEP:2003aa}, Tevatron~\cite{Carena:2004xs} and LHC~\cite{Amrith:2018yfb} 
  constraints which indicate $v_\Phi^2 \gg v^2$.

We now consider the $Z^\prime$ boson decay. 
Since we focus on the $Z^\prime$ boson mass in the multi-TeV range in this paper, 
  we neglect the SM fermion masses and for this case, the partial decay width formulas are listed in \cite{Das:2017flq}.
In the following analysis, we assume the RHNs are all heavier than the $Z^\prime$ boson, for simplicity. 
Since the SM Higgs doublet has the $U(1)_X$ charge $-\frac{1}{2} x_H$, 
  its gauge interactions leads to the trilinear coupling $Z^\prime-Z-h$. 
The relevant part of the Lagrangian is given by
\bea
\mathcal{L} &\supset& \Big|\Big\{-\frac{i}{2} g_Z Z_\mu -i g_X Z_\mu^\prime (-\frac{1}{2} x_H)\Big\}  \frac{1}{\sqrt{2}} (v+h)  \Big|^2 \nonumber \\
&=& \frac{1}{8} \Big(  g_Z^2  Z_\mu Z^\mu + g_X^2 x_H^2 Z^\prime_\mu {Z^\prime}^\mu -2 g_Z \Big( g_X x_H\Big) Z^\prime_\mu Z^\mu \Big) v^2 \Big(1+2\frac{h}{v}+\frac{h^2}{v^2}\Big) \nonumber\\
&\supset& -\frac{1}{2} g_Z (g_X x_H) vh Z^\mu Z^\prime_\mu =-M_Z \Big( g_X x_H\Big) h Z^\mu Z^\prime_\mu.
\label{Kin-scl}
\eea
As we expect, the trilinear coupling is proportional to $g_X x_H$. 
We then calculate the partial decay width of the  process $Z^\prime \to Zh$ to be
\bea
\Gamma(Z^\prime \to Z h) &=& \frac{ {g_X}^2 {x_H}^2}{48 \pi} M_{Z^\prime}
\sqrt{\lambda[a, b ,c]} 
\left(  \lambda[a, b, c] + 12 b  \right)  \simeq  \frac{ {g_X}^2 {x_H}^2}{48 \pi} M_{Z^\prime}, 
\label{zh}
\eea
where $\lambda[x, y, z]= x^2+y^2+z^2-2xy-2yz-2zx$, 
  and $a=1$, $b= \left( \frac{M_Z}{M_{Z^\prime}} \right)^2 \ll1 $ 
  and $c= \left( \frac{m_h}{M_{Z^\prime}} \right)^2 \ll1 $, respectively, 
  with the $Z$ boson mass $M_Z=91.2$ GeV and the Higgs boson mass $m_h=125$ GeV. 
  \textcolor{red}{The partial decay with of $Z^\prime$ into fermion-antifermion pairs can be given by
  \bea
  \Gamma(Z' \to \bar{f} f)
    &= N_C^{} \frac{M_{Z'}^{} g_{X}^2}{24 \pi} \left[ \left( q_{f_L^{}}^2 + q_{f_R^{}}^2 \right) \left( 1 - \frac{m_f^2}{M_{Z'}^2} \right) + 6 q_{f_L^{}}^{} q_{f_R^{}}^{} \frac{m_f^2}{M_{Z'}^2} \right]~,  
 \eea 
where $m_f$ is the mass of the fermion, $q_{L,R}$ are the corresponding $U(1)_X$ charges of the left and right handed fermions from Tab.~\ref{tab1}. }  
 
\begin{figure}
\begin{center}
\includegraphics[scale=0.2]{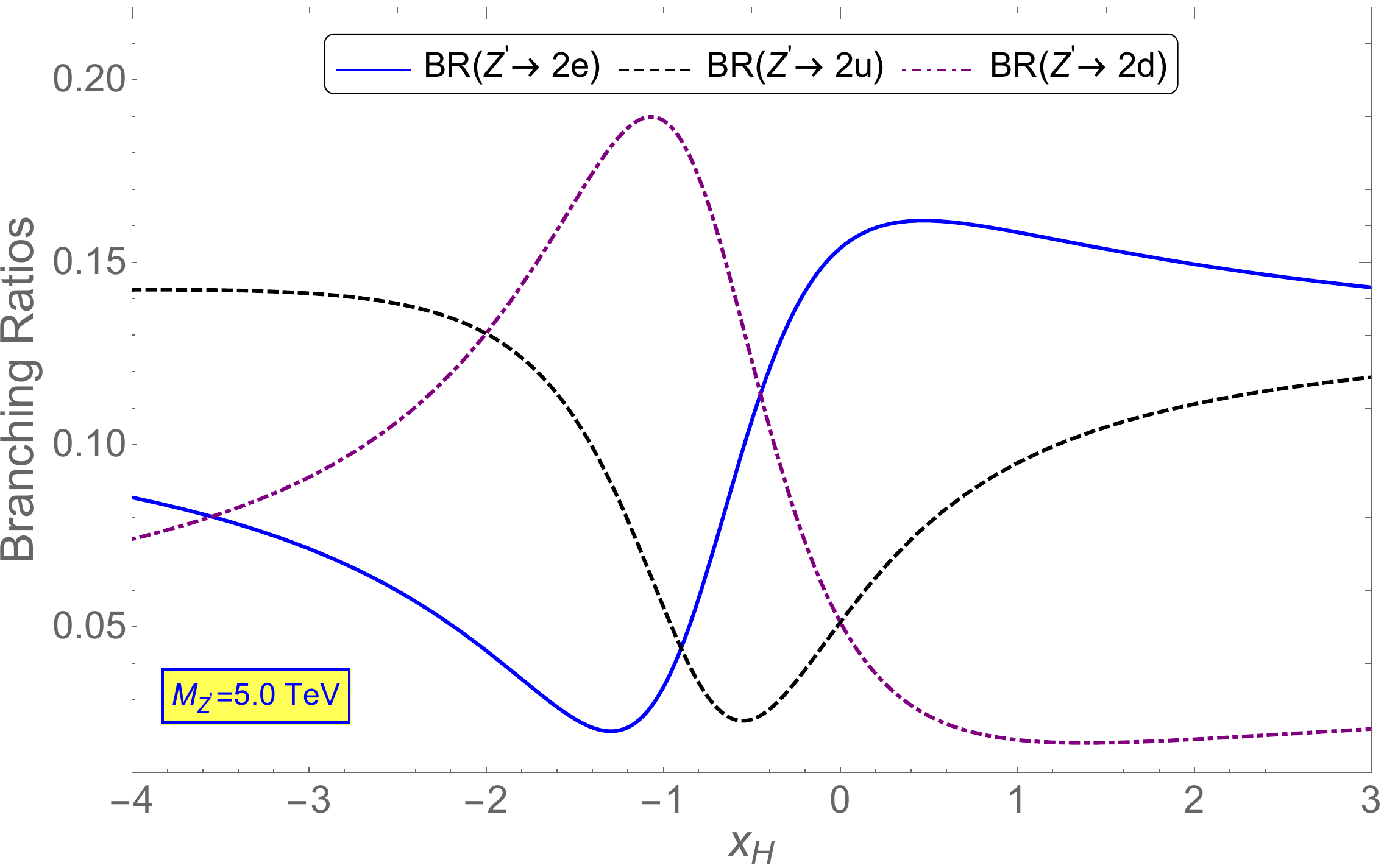} 
\includegraphics[scale=0.2]{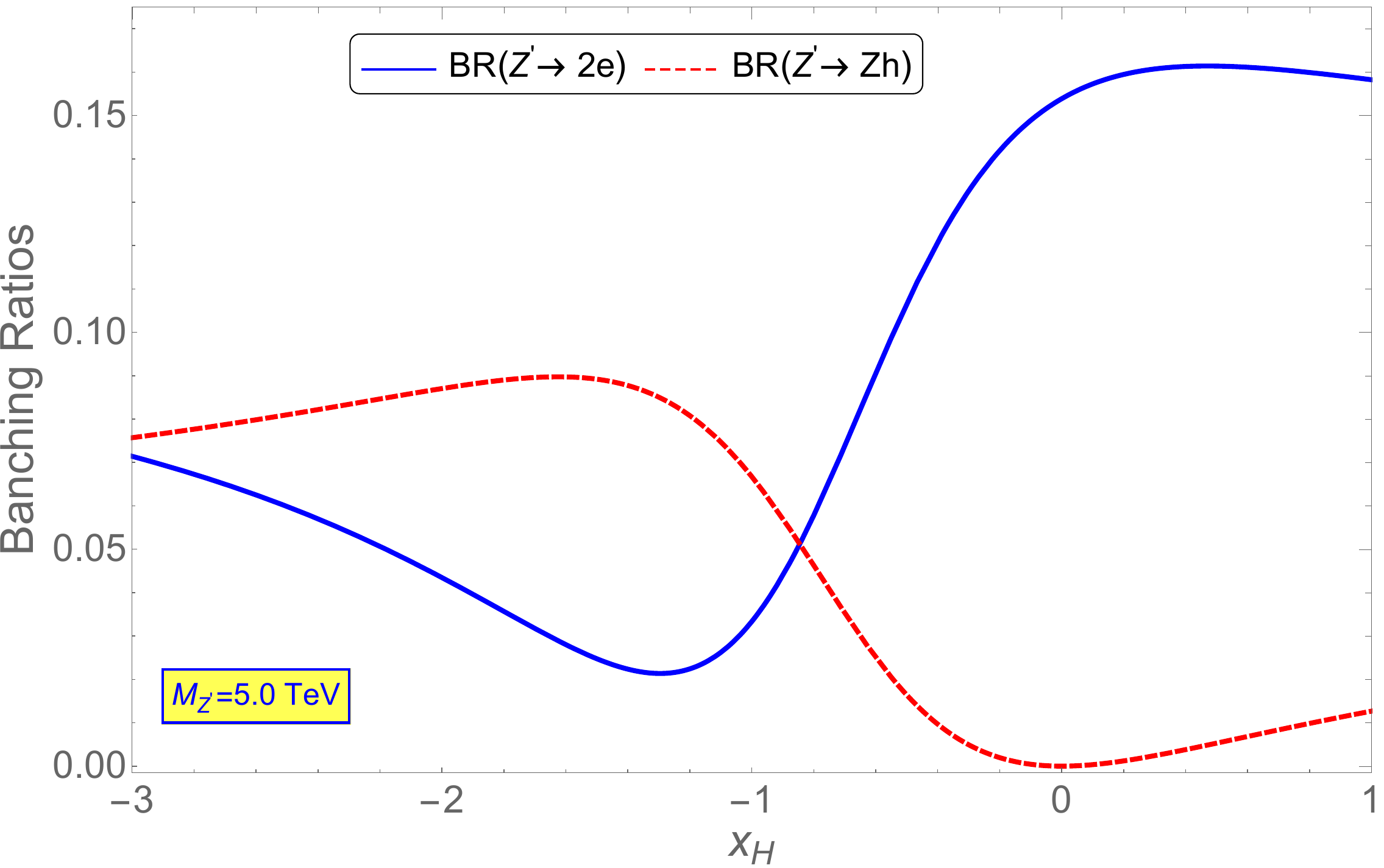}
\caption{Left Panel: Branching ratios of the $Z^\prime$ into di-electron (blue, solid), up-type di-quark (black, dashed) and down-type di-quark (purple, dot dashed) quarks as a function of $x_H$. 
Right Panel: Comparison between the branching rations, 
${\rm BR}(Z^\prime \to 2e)$ (red, dashed) and ${\rm BR}(Z^\prime \to  Zh)$ (blue, solid) 
as a function of $x_H$. 
We have fixed $M_{Z^\prime}=5$ TeV.}
\label{BRZpll1}
\end{center}
\end{figure}

\begin{figure}
\begin{center}
\includegraphics[scale=0.21]{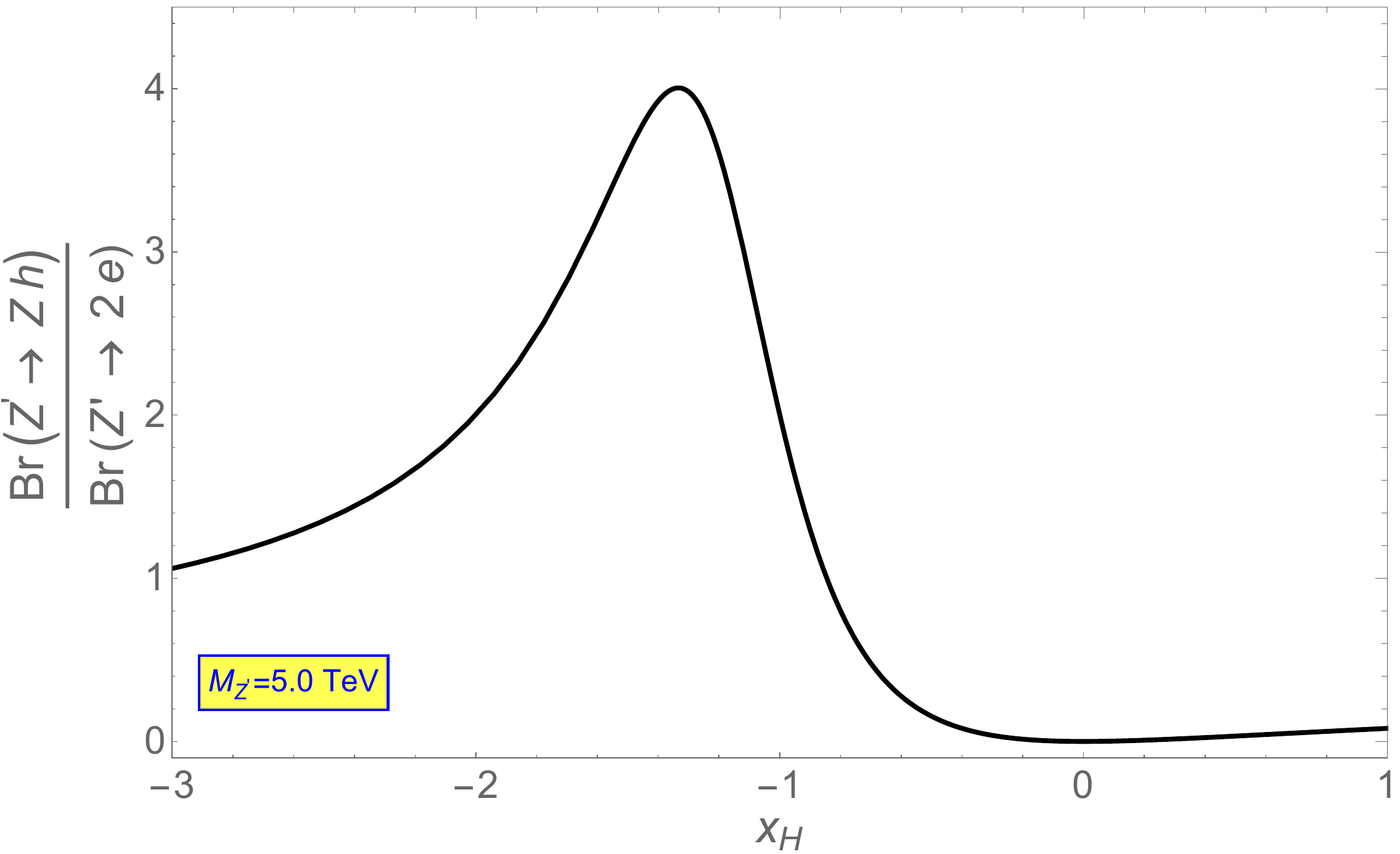}\\
\caption{Ratio of the branching ratios of $Z^\prime$ decaying to $Zh$ to $2e$ as a function of $x_H$ for $M_{Z^\prime} =5$ TeV.}
\label{BRZpll2}
\end{center}
\end{figure}

In the left panel of Fig.~\ref{BRZpll1}, we show the branching ratios to pairs of up-type quarks, down-type quarks and 
  charged leptons, respectively, as  a function of $x_H$, for $M_{Z^\prime}=5$ TeV. 
We can see the dramatic change of the branching ratios. 
In the right panel, we show the branching ratio to the $Z h$ mode as a function of $x_H$, 
  along with the dilepton mode. 
Since the most severe bound on $Z^\prime$ boson production at the LHC 
  is obtained from the final state dilepton search, 
  we are particularly interested in the ${\rm BR}(Z^\prime \to Zh)$ value compared with the ${\rm BR}(Z^\prime \to 2e)$ value. 
We can see that the $Z^\prime \to Zh$ mode dominates over the $Z^\prime \to 2e$ mode
  for $-3 \lesssim x_H \lesssim -0.8$.  
The ratio of the two branching rations is shown in Fig.~\ref{BRZpll2} as a function of $x_H$ for $M_{Z^\prime}=5$ TeV. 
The ratio exhibits its maximum at $x_H \simeq -\frac{4}{3}$, where ${\rm BR}(Z^\prime \to Zh) \simeq 4.01 \times  {\rm BR}(Z^\prime \to 2e)$. 
For $M_{Z^\prime}^2 \gg M_Z^2, m_h^2$, this ratio is approximately expressed as 
\bea
\frac{{\rm BR}(Z^\prime \to Zh)}{{\rm BR}(Z^\prime \to 2e)}
\simeq  \frac{x_H^2}{\frac{5}{2}x_H^2+6x_H+4}. 
\eea

To conclude this section, for readers convenience, 
  we give the Higgs boson coupling with a pair of $Z$ bosons in the SM
  from the $SU(2)_L \otimes U(1)_Y$ gauge interactions of the SM Higgs doublet.   
The relevant part of the Lagrangian is given by 
\bea
\mathcal{L} &\supset& \Big|-\frac{i}{2} g_z Z_\mu \frac{1}{\sqrt{2}} (v+h)\Big|^2 \nonumber \\
&=& \frac{g_z^2}{8} Z_\mu Z^\mu (v^2+2vh+h^2) \supset \frac{M_Z^2}{v} h  Z_\mu Z^\mu. 
\label{zzh}
\eea
Through the interactions in Eqs.~(\ref{zh}) and (\ref{zzh}) in the minimal $U(1)_X$ model, 
  the Higgs boson can be produced in association with a $Z$ boson 
  at the high energy colliders.

\section{$Zh$ production at the high energy colliders}
\label{Sec3}

\begin{figure}
\begin{center}
\includegraphics[scale=0.41]{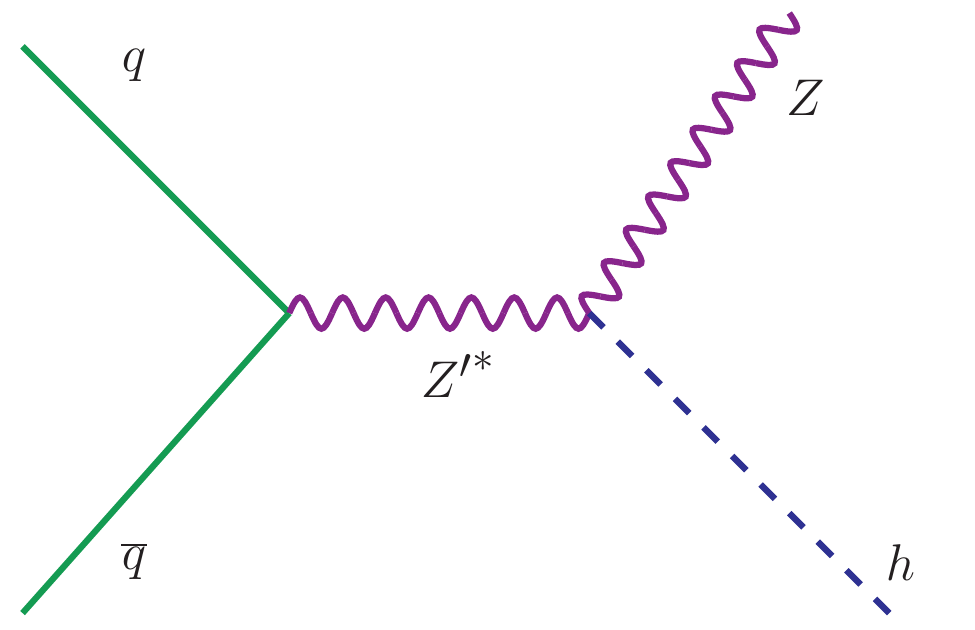} 
\includegraphics[scale=0.41]{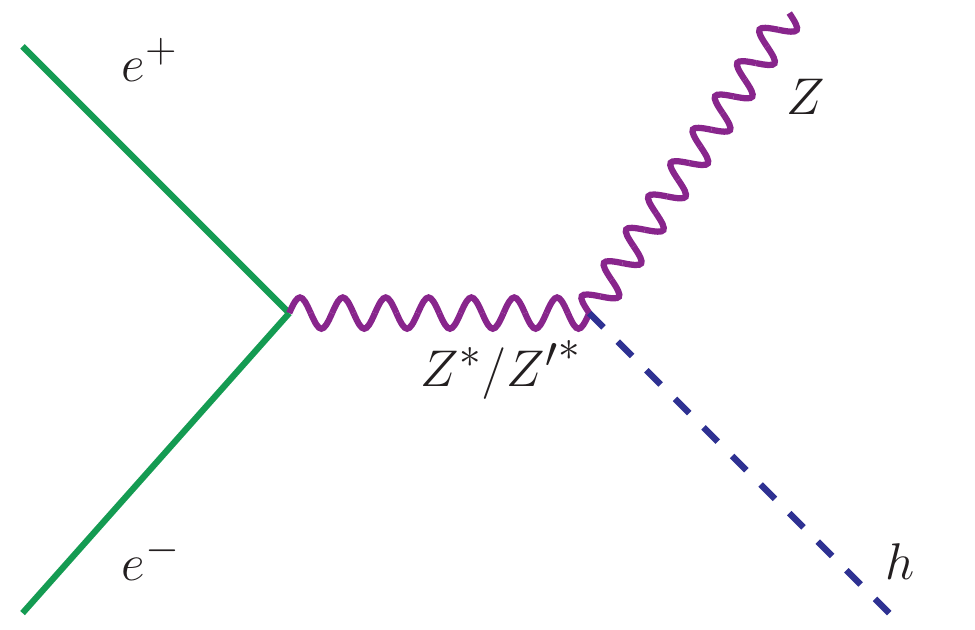}
\caption{Production mechanism of the SM Higgs boson in association with a $Z$ boson 
  at the proton-proton hadron collider (left) and the electron-positron collider (right). 
}
\label{FDZp}
\end{center}
\end{figure}

In this section, we study the $Z h$ production at the high energy colliders, 
 namely, the proton-proton collider and the electron-positron collider. 
At both colliders, the final state $Zh$ is produced through the $Z$ and $Z^\prime$ bosons in the $s$-channel. 
For the $Z^\prime$ boson with mass in the multi-TeV range, 
  there is an essential difference between the $Z^\prime$ boson phenomenologies at the two colliders. 
The energy of the proton-proton collider is high enough to produce the $Z^\prime$ boson resonance, 
  so that we focus on the $Z h$ production through the resonant $Z^\prime$ boson production 
  and its subsequent decay to $Z h$ (for this process, see the left Feynman diagram in Fig.~\ref{FDZp}).  
On the other hand, the energy of the electron-positron collider will not be high enough,  
  and the $Z^\prime$ boson effect on the $Z h$ production is mainly from 
  the interference between the $Z$ and $Z^\prime$ mediated processes 
  (see the right Feynman diagram in Fig.~\ref{FDZp}).

\subsection{Proton-proton collider}

\begin{figure}
\begin{center}
\includegraphics[scale=0.3]{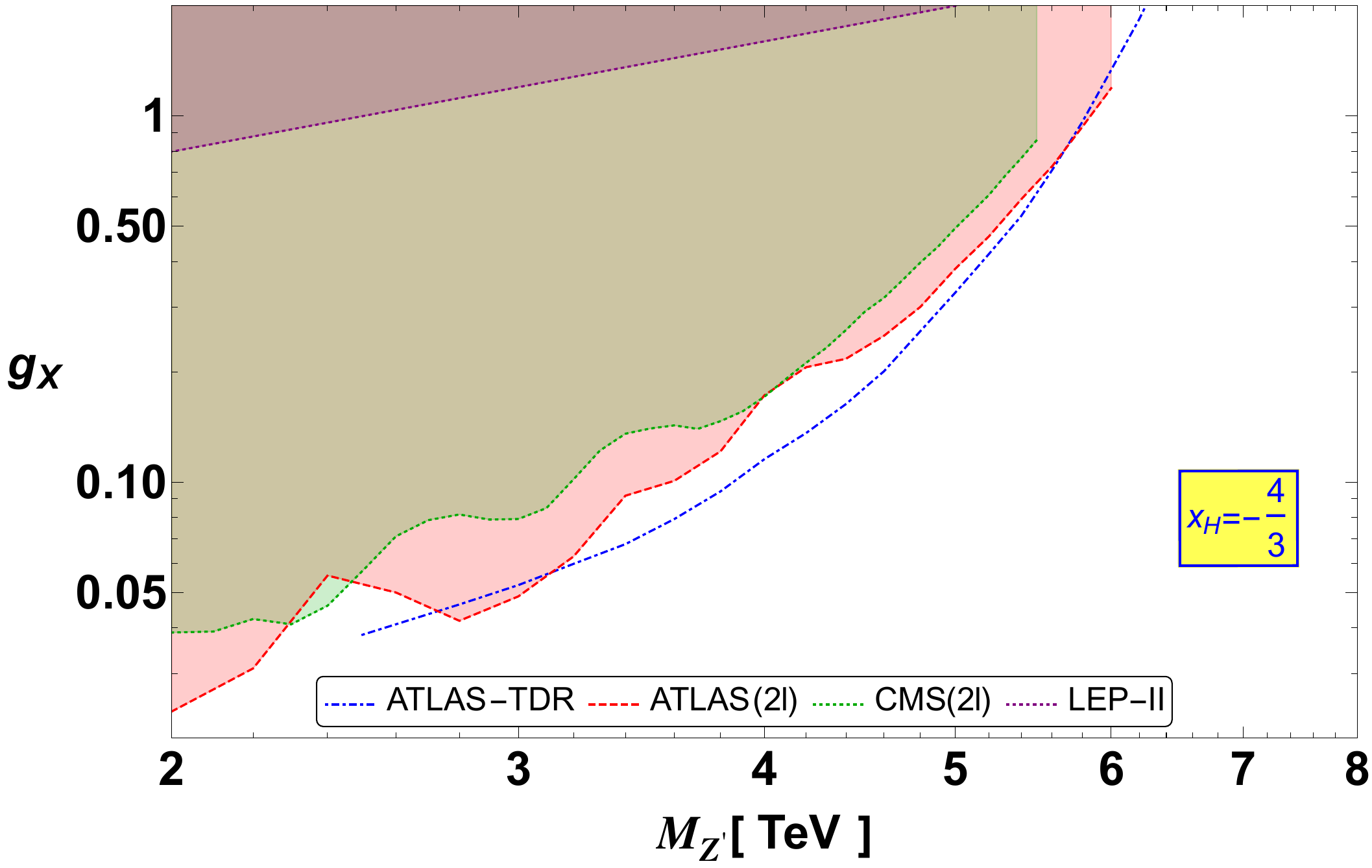} 
\caption{The prospective upper bound on $g_X$ for $M_{Z^{\prime}}=7.5$ TeV by the HL-LHC experiment 
  with a 3000 fb$^{-1}$  luminosity. }
\label{gX-MZp}
\end{center}
\end{figure}

We first consider the $Zh$ production at the proton-proton collider. 
As previously mentioned, we focus on the $Zh$ production 
  from the resonant $Z^\prime$ boson production and its subsequent decay to $Z h$.   
Since $g_X$ is constrained to be small from the LHC constraint, 
  the total $Z^\prime$ boson decay width is vary narrow. 
In tis case, we can use the narrow width approximation to evaluate the $Z^\prime$ boson 
  production cross section at the proton-proton collider.  
In this approximation, we express the total $Z^\prime$ boson production cross section by 
\bea 
   \sigma( pp \to Z^\prime) \ = \ 2 \sum_{q, \, \bar{q}} \int d x \int dy \, f_q(x, Q) \,  f_{\bar{q}} (y, Q) \, \hat{\sigma}(\hat{s}), 
 \label{Xsec1}  
\eea
  where$f_q$ ($ f_{\bar{q}}$) is the parton distribution function (PDF) for a quark (anti-quark), 
   $\hat{s}= x y s$ is the invariant mass squared of the colliding partons (quarks) 
   with a center-of-mass energy $\sqrt{s}$ of the proton-proton collider, 
   and the $Z^\prime$ boson production cross section at the parton level is expressed as
\bea
  \hat{\sigma}(\hat{s}) \ = \ \frac{4 \pi^2}{3} \frac{\Gamma(Z^\prime \to q \bar{q})}{m_{Z^\prime}} 
   \, \delta (\hat{s}-m_{Z^\prime}^2) .  
 \label{Xsec2}  
\eea             
Here, $\Gamma(Z^\prime \to q \bar{q})$ is the $Z^\prime$ boson partial decay width to $q \bar{q}$. 
For the decay modes to the up-type and down-type quarks are explicitly given by 
\bea
  \Gamma(Z^\prime \to u \bar{u}) &=&  \frac{g_X^2 }{288 \pi}  (8 + 20 x_H + 17 x_H^2) M_{Z^{\prime}}, \nonumber \\ 
   \Gamma(Z^\prime \to d \bar{d}) &=&  \frac{g_X^2 }{288 \pi}  (8 - 4 x_H + 5 x_H^2) M_{Z^{\prime}}. 
\eea
In our analysis, we employ \texttt{CTEQ6L} \cite{Pumplin:2002vw} for the PDF
  with a factorization scale $Q=M_{Z^\prime}$.  
We then simply express the $Zh$ production cross section by 
\bea
  \sigma( pp \to Z^\prime \to Zh) 
  \simeq \sigma( pp \to Z^\prime) \times {\rm BR}(Z^\prime \to Zh). 
\eea
As we have investigated in the previous section, ${\rm BR}(Z^\prime \to Zh)$ highly depends on $x_H$ and, in particular, it is enhanced for $x_H \simeq -\frac{4}{3}$ compared to ${\rm BR}(Z^\prime \to 2e)$. 
To estimate the constraints on the $U(1)_X$ coupling, we use the UFO model file\footnote{
The model file can be obtained from the link: https://feynrules.irmp.ucl.ac.be/wiki/GeneralU1no1/} 
of the general U(1) model \cite{Das:2021esm} in MadGraph \cite{Alwall:2011uj,Alwall:2014hca} 
and parton distribution function CTEQ6L \cite{Pumplin:2002vw} by fixing the factorization scale $\mu_F$ at the default MadGraph option. 
We simulate the dilepton production cross sections for a benchmark $g_{\rm Model}$ 
which includes the $Z^\prime$ contribution producing a cross section $\sigma_{\rm Model}$. 
%
We then compare our cross section with the observed dilepton cross sections from the ATLAS \cite{Aad:2019fac} 
at $139$ fb$^{-1}$ luminosity studied for the sequential Standard Model (SSM) \cite{Langacker:2008yv} at $\frac{\Gamma}{m}=3\%$ 
by using
\bea
g^\prime = \sqrt{g_{\rm{Model}}^2 \Big(\frac{\sigma_{\rm{ATLAS}}^{\rm{Observed}}}{\sigma_{\rm{Model}}}\Big)} .
\label{gp}
\eea
Since the constraints from the dijet process is weaker than the dilepton bounds, we do not consider them. 
The bound obtained from the LEP-II considering $M_{Z^\prime} > \sqrt{s}$ we find $\frac{M_{Z^\prime}}{g_X} > 2.51$ TeV \cite{Okada:2016tci}. 
The limits are shown in Fig.~\ref{gX-MZp}, \textcolor{red}{where we find that $g_X=0.3$ and 1 for $M_{Z^\prime}=5$ TeV and 7.5 TeV respectively.}

\begin{figure}
\begin{center}
\includegraphics[scale=0.18]{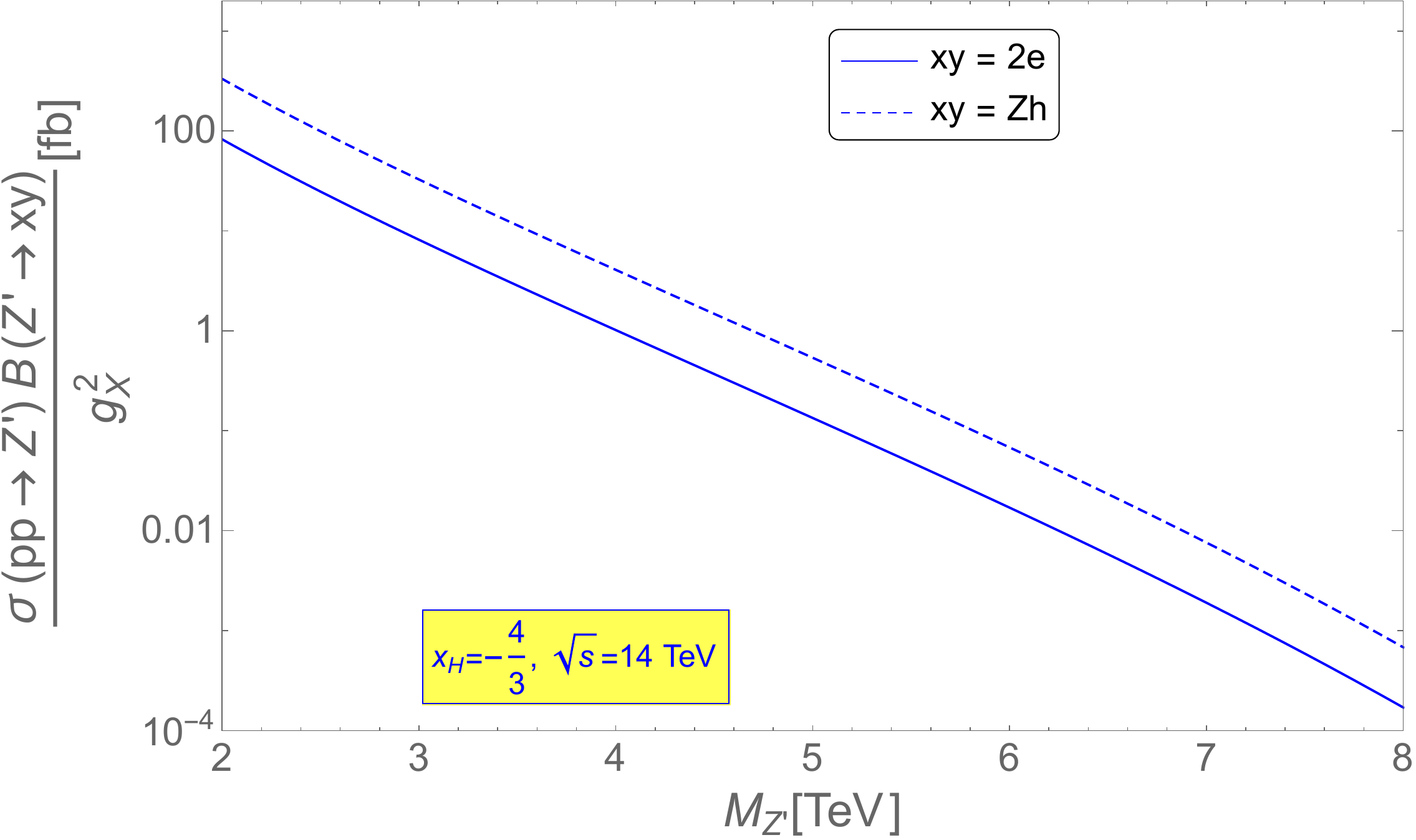}
\includegraphics[scale=0.18]{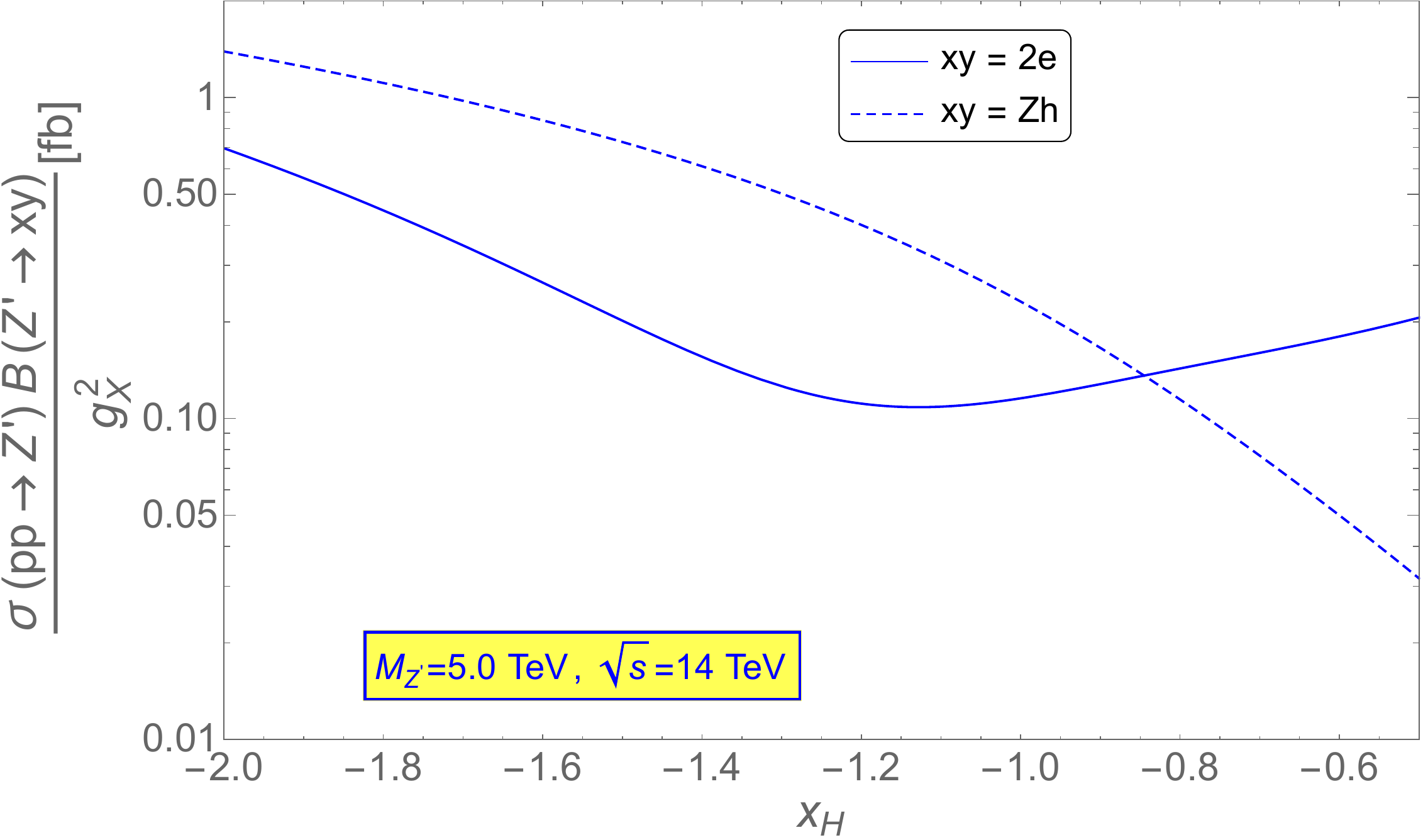}\\
\includegraphics[scale=0.18]{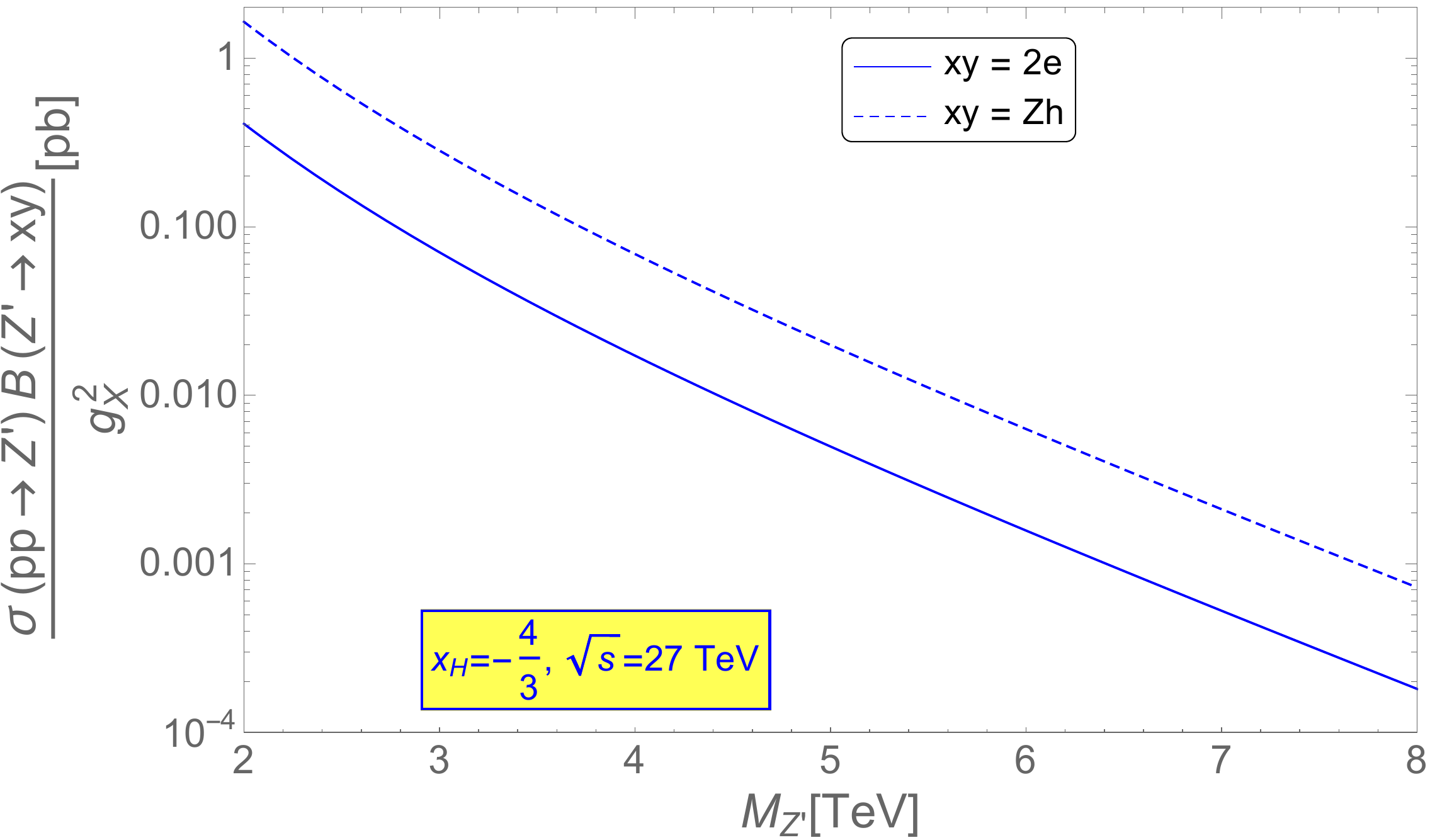}
\includegraphics[scale=0.18]{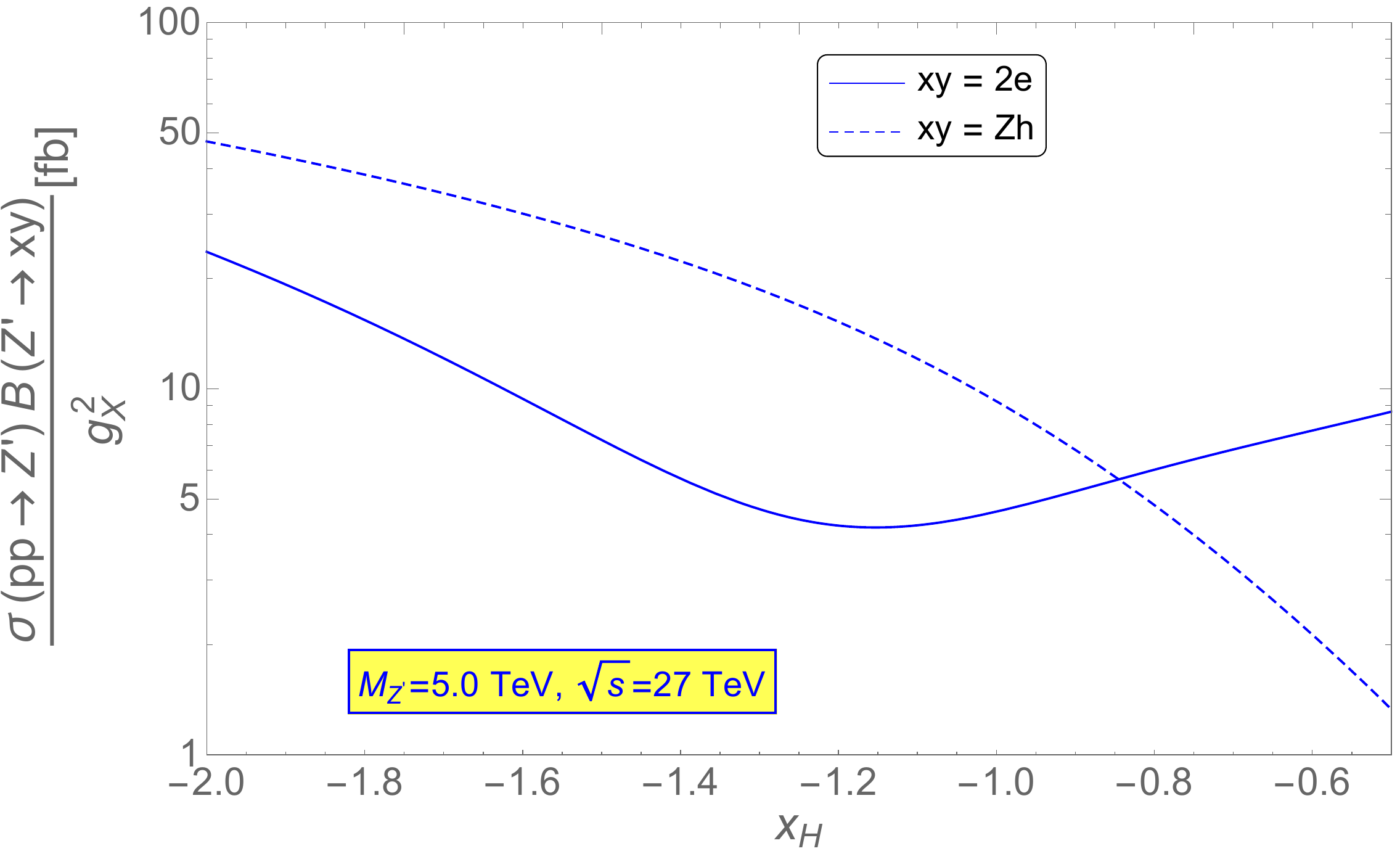}
\caption{
The $Zh$ production cross section (blue, dashed) from the $Z^\prime$ boson decay 
  as a function of $Z^\prime$ mass $(M_{Z^\prime})$ (left column) and $x_H$ (right column). 
The corresponding dilepton production cross sections (blue, solid) are shown for comparison. 
}
\label{Zpll3}
\end{center}
\end{figure}

We are now ready to show the results for our production cross section calculations. 
In addition to the LHC energy of $\sqrt{s}=14$ TeV for Run 3 and High-Luminosity upgrade, 
  we have examined various center-of-mass energies 
  for the future hadron collider projects, such as $\sqrt{s}=27$ TeV and $100$ TeV respectively. 
The dilepton production cross section as a function of $M_{Z^\prime}$ for three different $x_H$ values and various $\sqrt{s}$ values. 
Since the cross section is proportional to ${g_X}^2$ in the narrow width approximation, we have shown the cross section normalized by ${g_X}^2$. 
Simultaneously we consider the $Zh$ production from the narrow width approximation and compare with the dilepton production cross section in Fig.~\ref{Zpll3}.
In the left column we compare both cross sections. In the left column we calculate the cross sections for fixed $x_H$ and center of mass energy varying $M_{Z^\prime}$.
In the right column we show the cross sections for fixed center of mass energy and $M_{Z^\prime}=5.0$ TeV. 

\subsection{Electron-positron collider}

Next we consider the $Zh$ production at the electron-positron collider. 
Since the $Z^\prime$ boson is too heavy to be produced at the electron-positron collider, 
  the $Z^\prime$ boson mediated process affects the $Zh$ production
  through the interference with the SM $Zh$ production through the $Z$ boson. 
  The interaction Lagrangian between the electron and $Z$ boson is given by
\bea
\mathcal{L}_{int}^Z =  g_Z \overline{e} \gamma^{\mu} \Big(C_V+C_A \gamma_5\Big) e Z_\mu ,
\eea
where $C_V= g_Z \left(-\frac{1}{4}+ \sin^2\theta_W \right)$ and $C_A= \frac{1}{4} g_Z$ .
The interaction between $Z^\prime$ and the electron is given by
\bea
\mathcal{L}_{int}^{Z^\prime}= \overline{e} \gamma^\mu \Big(C_V^\prime+ C_A^\prime \gamma_5\Big) e Z_\mu^\prime
\eea
where $C_V^\prime= g_X \left(-\frac{3}{4}x_H-1\right)$ and $C_A^\prime = g_X \left(-\frac{1}{4} x_H \right)$ .
We then calculate the differential cross section of the $Zh$ production at the electron-positron collider to be
\bea
\frac{d\sigma^{ee}}{d \cos\theta} &=& \frac{1}{32\pi} \sqrt{\frac{E_Z^2-M_Z^2}{s}} \Big[ \Big|C_Z\Big|^2 \Big(C_V^2+C_A^2\Big)+ |C_Z^\prime|^2 \Big({C^{\prime}_{V}}^{2}+ {C^{\prime}_{A}}^{2}\Big) \nonumber \\
&+& \Big( C_Z^\ast C_Z^\prime+ C_Z {C_{Z}^\prime}^\ast\Big)  \Big( C_V C_V^\prime+ C_A {C_{A}^\prime}\Big)\Big] \times \Big\{1+\cos^2\theta+\frac{E_Z^2}{M_Z^2} \Big(1-\cos^2\theta\Big)\Big\}, 
\eea
where $E_Z = \frac{s+M_Z^2 - m_h^2}{2 \sqrt{s}}$, $C_Z=2 \Big(\frac{M_Z^2}{v}\Big) \frac{1}{s-M_Z^2+i\Gamma_Z M_Z}$ and $C_Z^\prime= \frac{-M_Z g_X x_H}{s-{M_Z^\prime}^2+i\Gamma_{Z^\prime}M_{Z^\prime}}$, respectively.
\begin{figure}
\begin{center}
\includegraphics[scale=0.29]{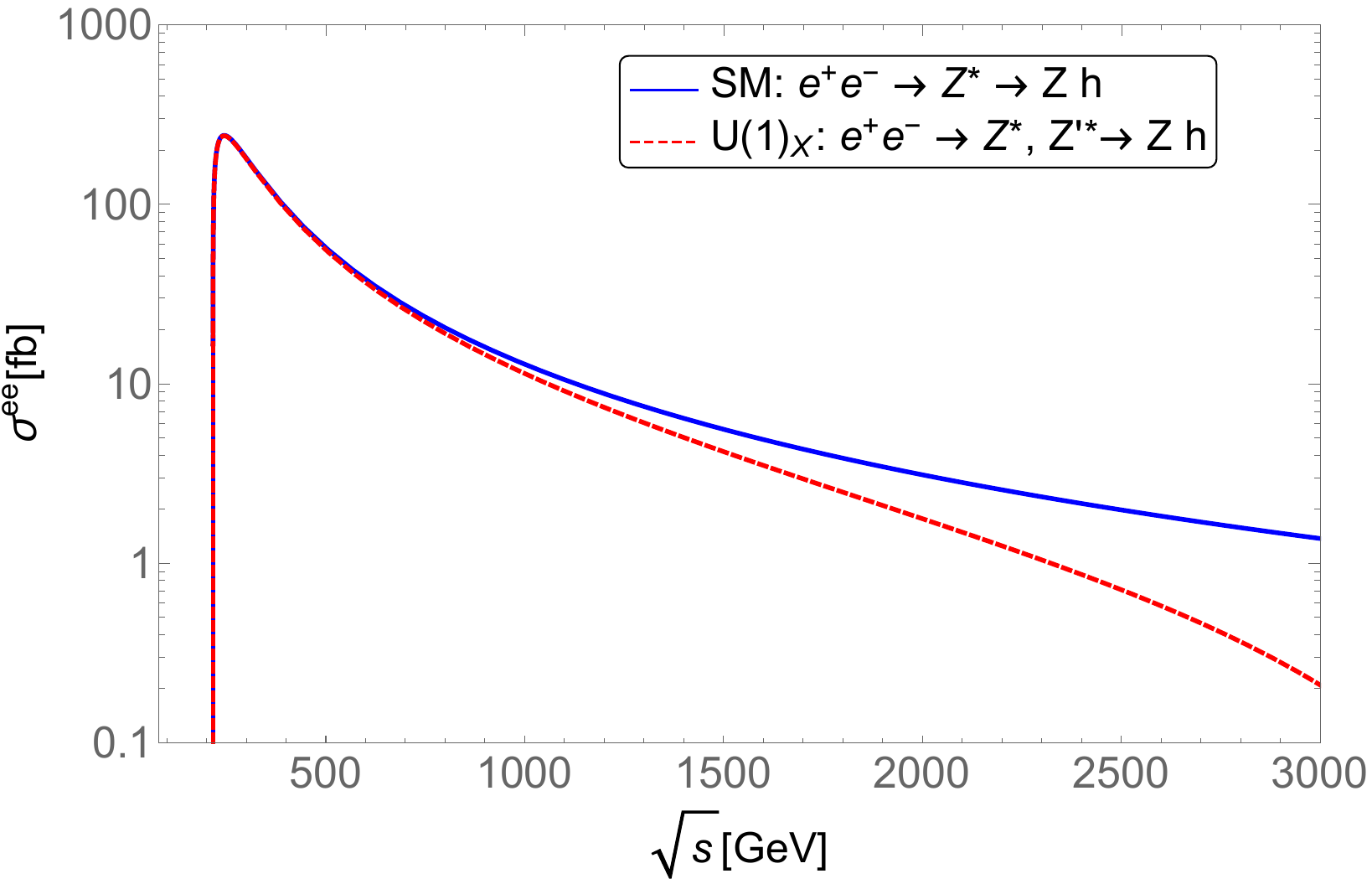}
\includegraphics[scale=0.194]{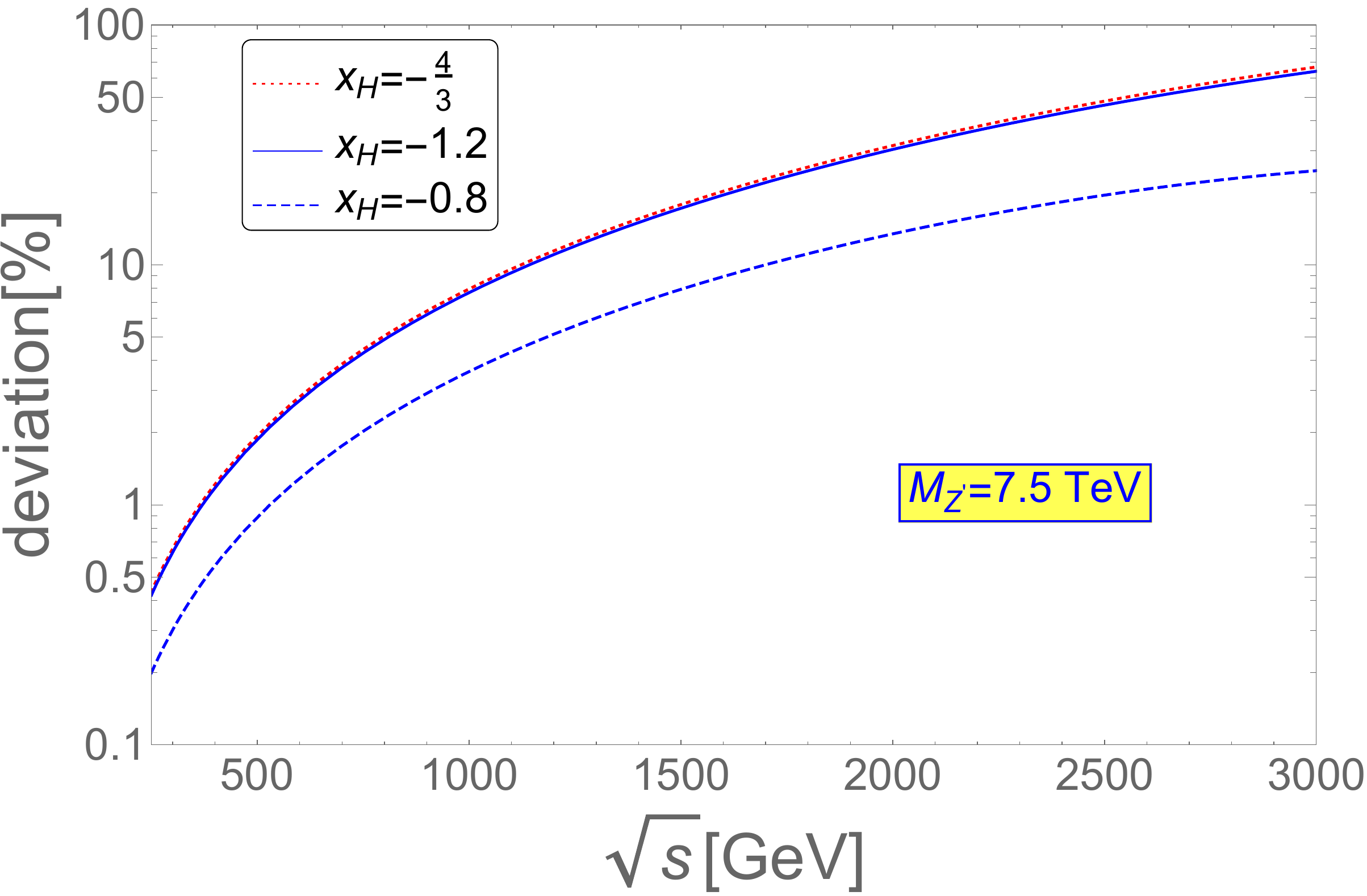}
\caption{Left panel: $Zh$ production cross section at the electron-positron collider in the SM (blue, solid) and the minimal $U(1)_X$ model (red, dashed) as a function of the center-of-mass energy $(\sqrt{s})$ of the collider. The interference between the $Z$ and $Z^\prime$ mediated processes is involved in the $U(1)_X$ case. Right panel: The deviation of the $Zh$ production cross section in the $U(1)_X$ model at the electron-positron collider as a function of the center-of-mass energy for fixed $x_H=-\frac{4}{3}$ (red, dotted) $-1.2$ (blue, solid) and $-0.8$ (blue, dashed)Here, we have fixed $M_{Z^\prime}=7.5$ TeV and $x_H=-\frac{4}{3}$.}
\label{Zpll}
\end{center}
\end{figure}
\begin{figure}
\begin{center}
\includegraphics[scale=0.2]{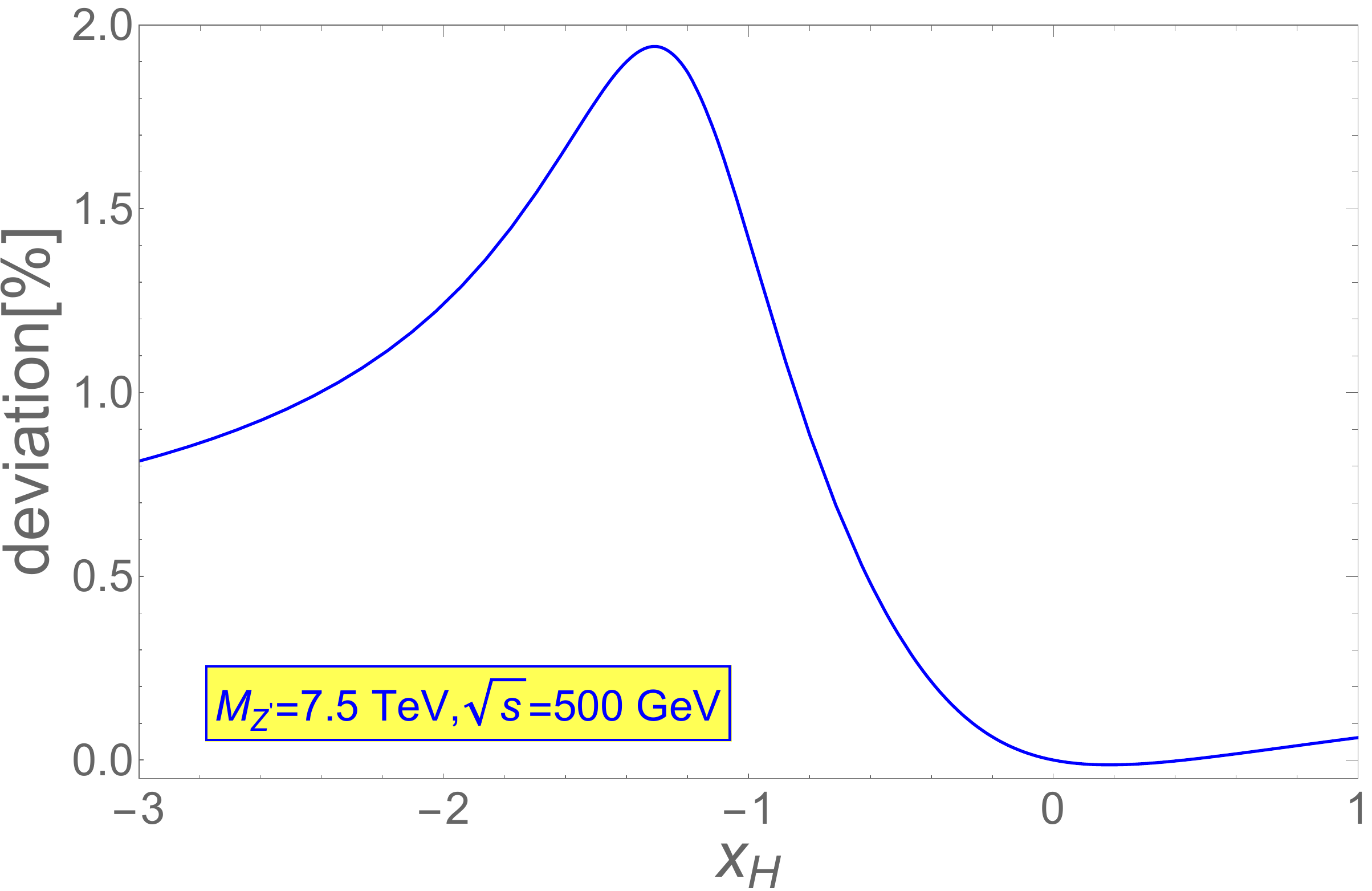} 
\includegraphics[scale=0.2]{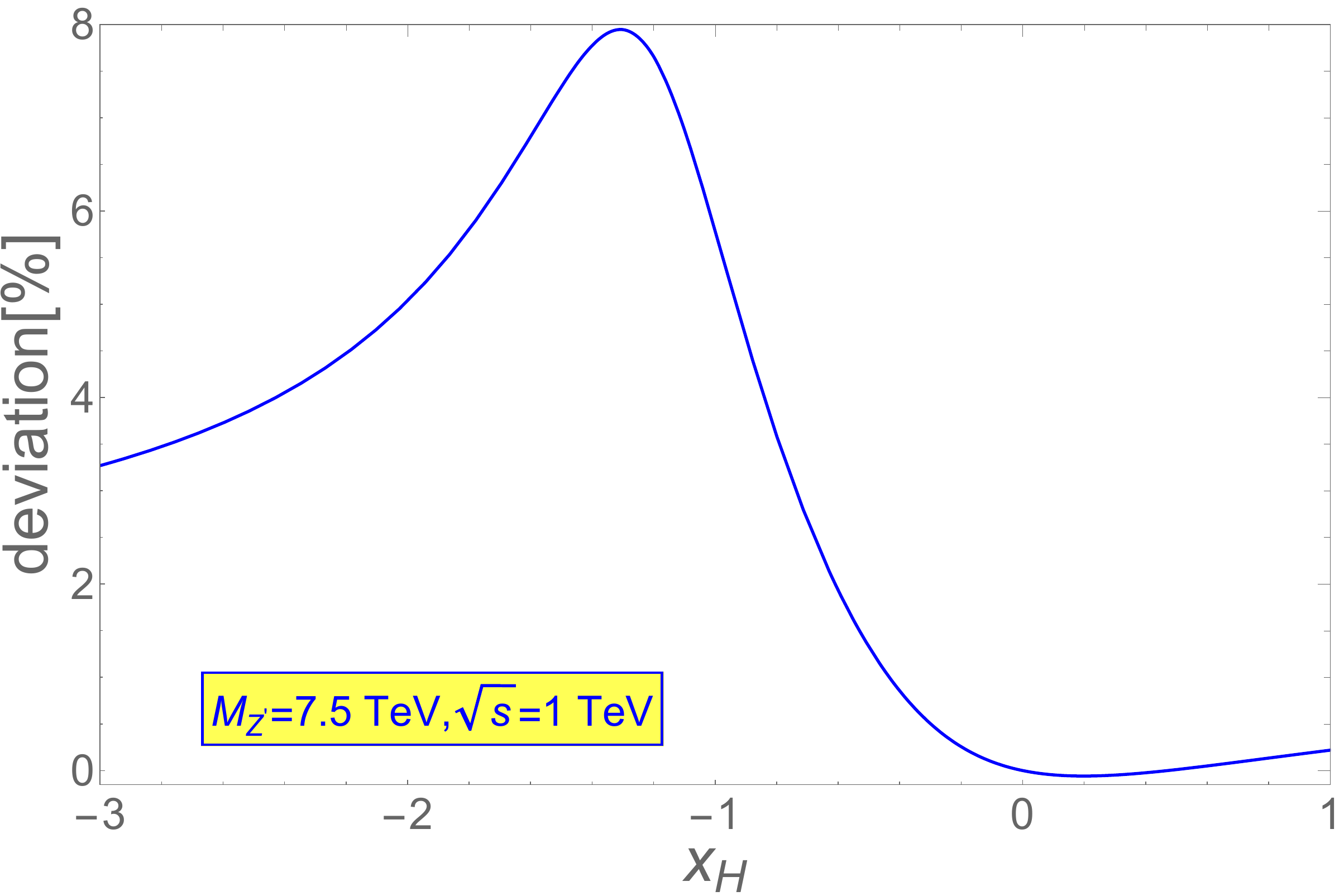}\\
\includegraphics[scale=0.2]{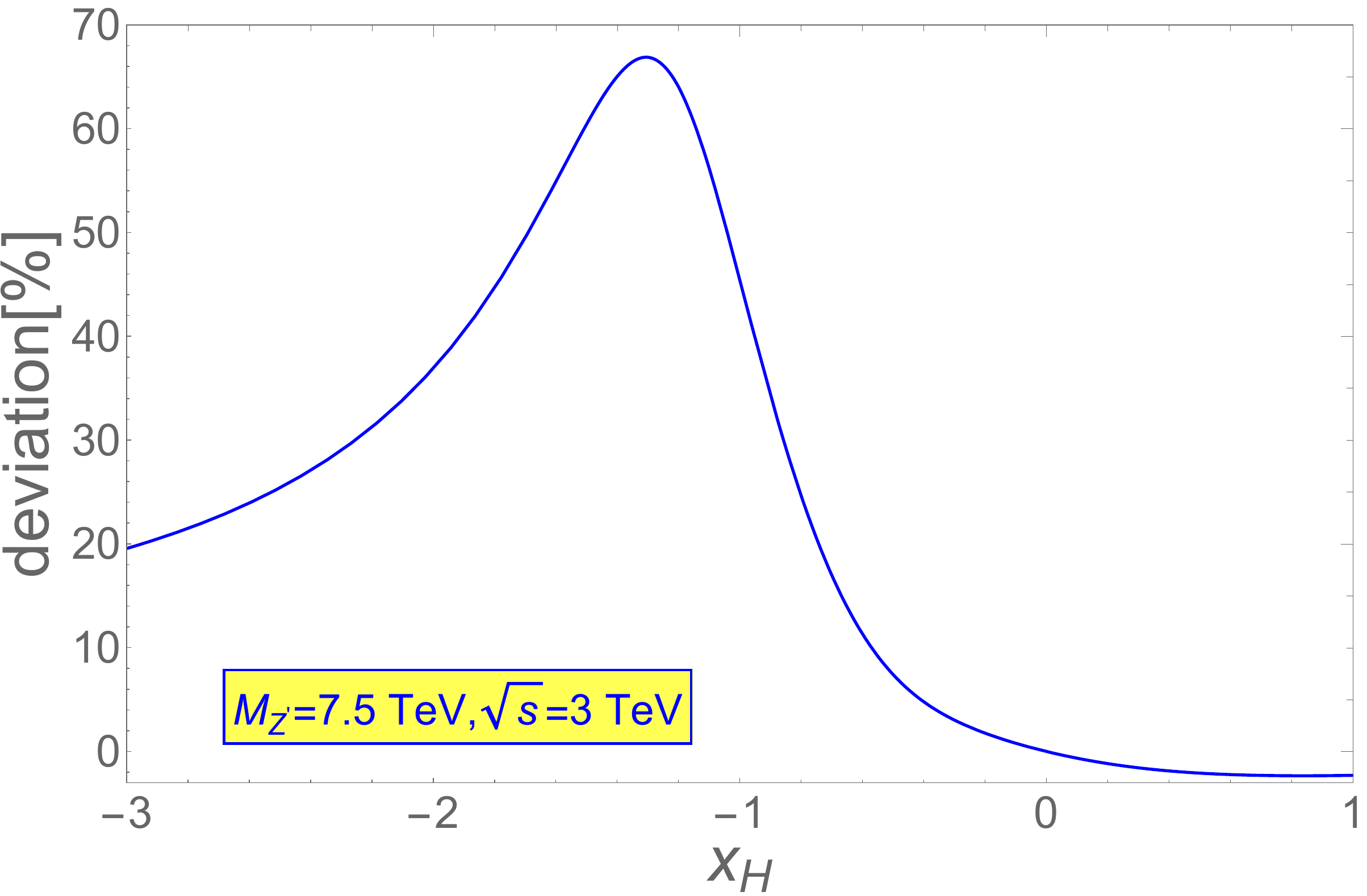}
\caption{The deviation of the $Zh$ production cross section in the $U(1)_X$ model form the SM for different $x_H$ values 
 and center-of-mass energy for the electron-positron collider, $\sqrt{s}=500$ GeV (top, left), $1$ TeV (top, right) and $3$ TeV (bottom), respectively. 
 We have fixed $M_{Z^\prime}=7.5$ TeV for $x_H=-\frac{4}{3}$.}
\label{dev2}
\end{center}
\end{figure}
The $Zh$ production cross section deviates from its SM prediction 
   by the contribution from the $Z^\prime$ boson mediated process. 
To evaluate this deviation which may be observed at the future electron-positron collider,  
   we calculate the $Zh$ production cross section by setting the model parameters 
   from the HL-LHC prospective bound, 
   namely, $M_{Z^\prime}=7.5$ TeV and $g_X$ can be found from Fig~\ref{gX-MZp}.   
In Fig.~\ref{Zpll}, we show the $Zh$ production cross section for $x_H=-\frac{4}{3}$ as a function of 
   the center-of-mass energy $\sqrt{s}$ of the electron-positron collider (blue, solid)  
   along with the SM prediction in the limit of $M_{Z^\prime} \to \infty$ (red, dashed). 
The interference between the $Z$ and $Z^\prime$ boson mediated processes 
   reduces the $Zh$ production cross section from its SM prediction. 
As expected, as the collider energy is raised, the deviation becomes larger. 
Let us define the deviation of the $Zh$ cross section of the minimal $U(1)_X$ model ($\sigma_{U(1)_X}$)
  from the SM predication ($\sigma_{SM}$) by 
\bea
Deviation[\%] = \left|1-\frac{\sigma_{U(1)_X}}{\sigma_{SM}} \right| \times 100 \, \%.
\label{devilc}
\eea
For three different values of $x_H$, we show in Fig.~\ref{Zpll} the deviation as a function of the collider energy.  
With its very clean environment, the electron-positron collider allows us to measure the $Zh$ production cross section 
  very precisely, say, an ${\cal O}(1\, \%)$ level of precision. 
In Fig.~\ref{Zpll}, we can see that for $x_H=-\frac{4}{3}$ the deviation from the SM cross section can be tested 
  at a 1 TeV electron-positron collider. 
In Fig.~\ref{dev2}, we show the deviations as a function of $x_H$ for 
  three different collider energies. 
  
\begin{figure}
\begin{center}
\includegraphics[scale=0.25]{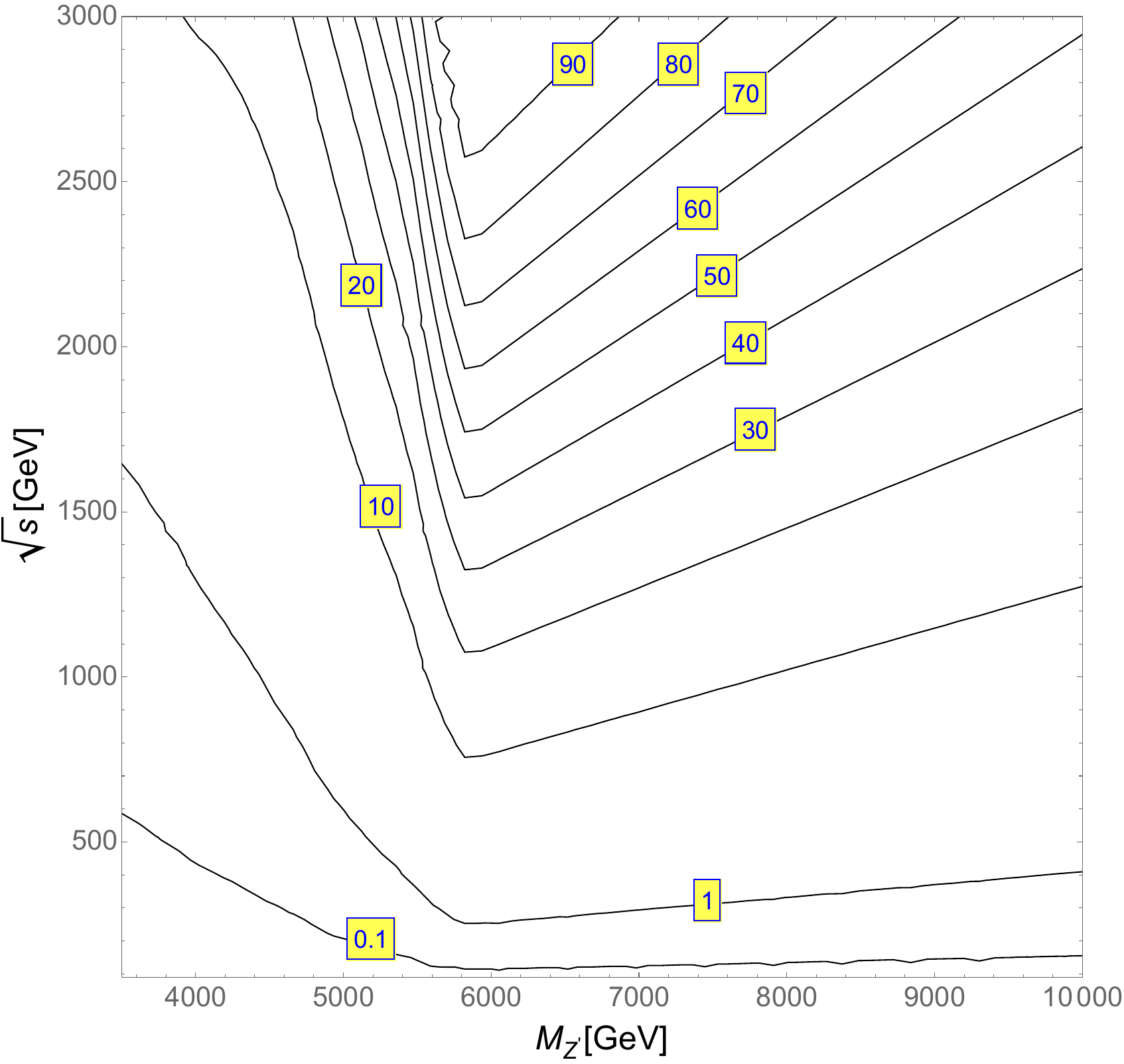} 
\includegraphics[scale=0.25]{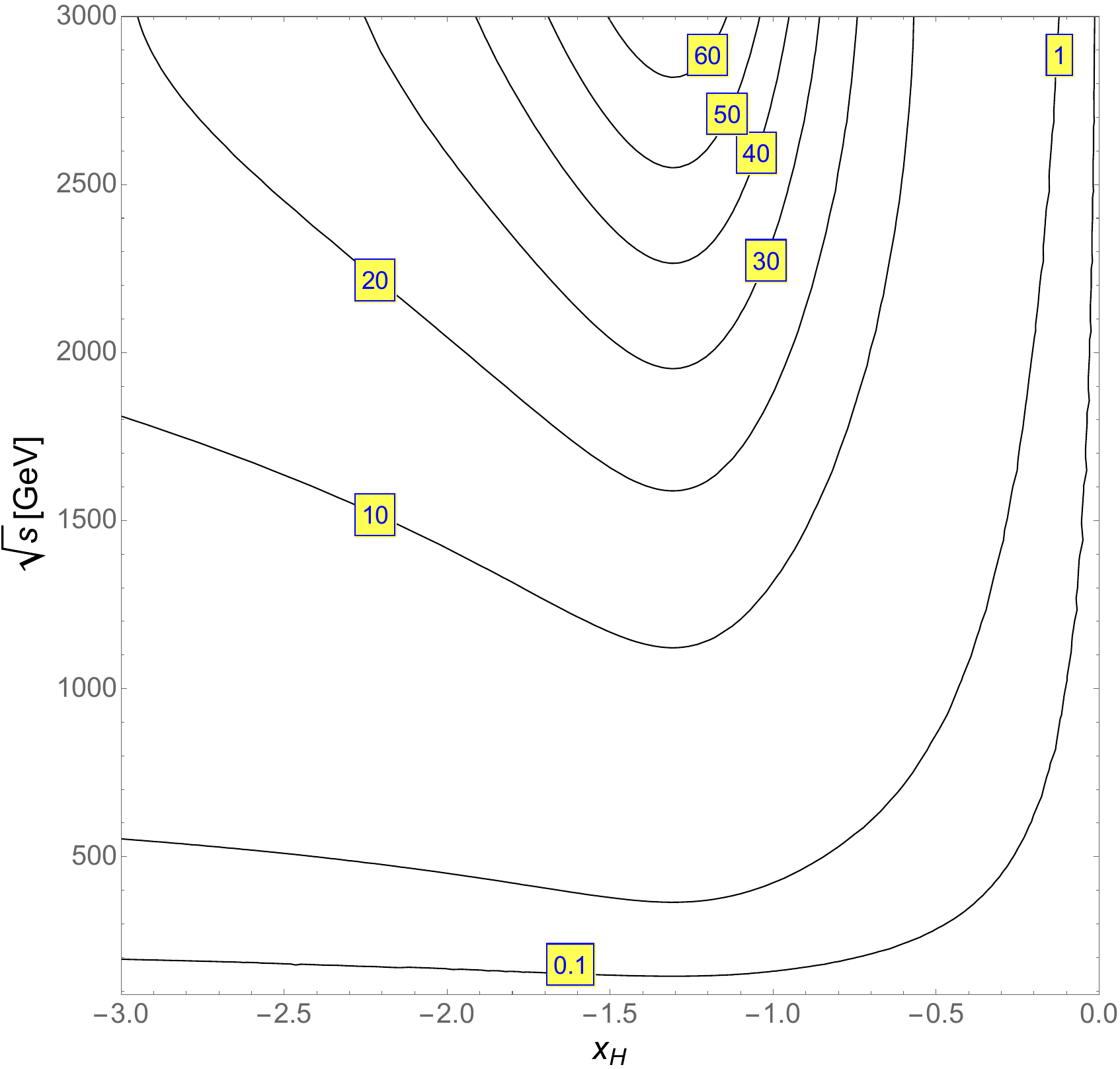}
\caption{Contours for deviations of the $Zh$ production cross section in the $U(1)_X$ model form the SM for different center of mass energies $(\sqrt{s})$ varying $M_{Z^\prime}$ at $x_H=-\frac{4}{3}$ (left panel) and varying $x_H$ at $M_{Z^\prime}=7.5$ TeV (right panel) respectively.}
\label{dev3}
\end{center}
\end{figure}

We calculate the deviations of the  $Zh$ production cross section at the electron positron collider from Eq.~(\ref{devilc}).
The deviations for different $\sqrt{s}$ and $M_{Z^\prime}$ are shown in the left panel of Fig.~\ref{dev3} for $x_H=-\frac{4}{3}$.
That for different $x_H$ fixing $M_{Z^\prime}$ at 7.5 TeV are shown in the right panel of  Fig.~\ref{dev3}.
Different contours for deviations in units of \% are shown in yellow. 
We find that a deviation in the cross section is around or below $10\%$ for $\sqrt{s}=1$ TeV and $M_{Z^\prime}=$ 7.5 TeV. 
The deviation decreases to the level of $1\%$ at $\sqrt{s}=250$ GeV.  
As is expected, the deviation in $Zh$ production cross section increases as the center of mass energy $\sqrt{s}$ is raised.
From Fig.~7 in Ref.~\cite{LCCPhysicsWorkingGroup:2019fvj}, we find that the $Z-h$ coupling could be measured with an accuracy of $0.25\%$ 
by combining the prospective High-Luminosity LHC results with  the ILC at 250 GeV, 500 GeV and 1 TeV results. 
Thus, we expect that a deviation of ${\cal O}(0.5\%)$ can be observed at the ILC, and a wide range of the parameter region
shown in Fig.~\ref{dev3} can be covered by the ILC. 
 

\section{Conclusion}
\label{Sec4}

The minimal $U(1)_X$ model is a simple and well-motivated BSM 
  which provides the origin of the neutrino mass 
  by the seesaw mechanism with three RHNs, which are necessary 
  for cancelling the $U(1)_X$ related anomalies. 
The $U(1)_X$ gauge boson $Z^\prime$ has been searched by the LHC experiments
  with the final state dilepton and dijets. 
In this paper, we have noticed that due to the $U(1)_X$ charge $-x_H/2$ of the SM Higgs doublet in the minimal $U(1)_X$ model, 
  the SM Higgs boson has a trilinear coupling of $Z^\prime-Z-h$ 
  which affects the associated Higgs boson production with a $Z$ boson at high energy colliders. 
We have calculated the $Zh$ production cross section at high energy proton-proton and electron-positron colliders
  and have shown interesting effects of the new $Z^\prime$ boson mediated process. 
The $Z^\prime$ boson contribution dramatically changes as a function of $x_H$, 
   and we have found that $x_H \simeq -\frac{4}{3}$ exhibits the maximum effect on the $Zh$ production. 
The precise measurement of the $Zh$ production cross section at the future high energy collider experiments 
   can reveal the $Z^\prime$ boson effect, 
   independently of the $Z^\prime$ boson search with the dilepton and dijet. 
\section*{Acknowledgement}
The work was supported in part by Japan Society for the Promotion of Science (JSPS), 
  Grant-in-Aid for Scientific Research, No.~18F18321 (A.D.) and
  by the United States Department of Energy grant DE-SC0012447 (N.O.).


\end{document}